
b
\magnification=1200
\def\a{\alpha}
\def\s{\sin^2\theta_w(M_w)}
\def\o{\otimes}
\def\y{\theta}
\def\c{\cdot}
\vskip 1cm
\baselineskip=27pt
\noindent
{\bf MAJORANA MASSES FOR NEUTRINOS IN SO(10)}
\vskip 3pt
\noindent
in Proceedings of the Second International Workshop on Neutrino Telescopes,
Venice 1990
\vskip 24pt
\noindent
{\bf G. Amelino-Camelia, F. Buccella and L. Rosa}
\vskip 3pt
\noindent
ISTITUTO DI FISICA TEORICA di NAPOLI
\vskip 48pt
The experimental values of $\alpha(M_w), \sin^2\theta_w(M_w)$\ and
$\alpha_s(M_w)^{[1]}$:
$$\eqalignno{\alpha(M_w)&=1/128\cr \sin^2\theta_w(M_w)&=0.228\pm.004 &(1)\cr
\alpha_s(M_w)&=0.107^{+.013}_{-.009}\cr }$$
are such that the gauge coupling $\a_3(t),\a_2(t)$\ and $\a_1(t)$\ of the
standard group $G=SU(3)_c\o SU(2)_L\o U(1)_Y$\ evolve as shown in fig.1.

Consequently they cross at three different scales .
In unified models with gauge group $SO(10)^{[2]}$\ it is not possible to
break SO(10) directly to G with only one of the smallest irreducible
representations for the Higgs scalars . In fact, as we can see in table 1, all
the singlets in these representations have symmetry group larger than
G$^{[3]}$.
So, we expect at least two energy scales for the spontaneous symmetry breaking
such that at the highest scale $(M_X)$\ SO(10) breaks down to $G'$\ and then,
at $M_R$, $G'$\ breaks to the standard group G.
If $G'$\ contains $SU(2)_R$\ and/or $SU(4)_{PS}$\ (in SO(10) $Y=T_{3R}+{B-L
\over2}$) we may obtain the unification of the $\a_i$\ with an appropriate
choice of $M_R$\ and $M_X$\ . We have the following possibilities for
$G^{'[4]}$:
\centerline{TABLE 1}
\settabs  2\columns
\+ $G'$   &{\ \ \ }Highest VEV in the\cr
\+ $SU(4)_{PS}\o SU(2)_L\o SU(2)_R\times D$    &{\ \ \ }54\cr
\+ $SU(4)_{PS}\o SU(2)_L\o SU(2)_R$ &   ${\ \ \ }\Phi_T=A_{78910}\in210$\cr
\+ $SU(3)_c\o SU(2)_L\o SU(2)_R\o U(1)_{B-L}\times D$ &${\ \ \ }\Phi_L={1\over
\sqrt3}(A_{1234}+A_{3456}+A_{1256})\in210$\cr
\+ $SU(3)_c\o SU(2)_L\o SU(2)_R\o U(1)_{B-L}$ & ${\ \ \ }\cos\theta\Phi_L+
\sin\theta\Phi_T\in210$\cr
\+  &  ${\ \ \ }(\sin2\theta\neq0)$\cr
\vskip 0.5cm
In all these cases the breaking scale $M_X$\ is not higher than the scale
at which $\a_2$\ and $a_3$\ joint in fig.1 . In fact, if $G'$\ contains
$SU(4)_{PS}$, $\a_3$, alias $\a_4$, decreases faster above $M_R$\
and meets $\a_2$\ earlier . If $G'$\ contains D$^{[5]}$, the left-right
symmetry
at the highest scale implies the existence of scalars with non trivial
properties under $SU(2)_L$\ with masses $\sim M_R$\ (this because it is
necessary the existence of scalars with masses $\sim M_R$\ and non trivial
properties under $SU(2)_R$\ to break this symmetry at $M_R$).
Because of the contributions of these scalars, $\a_{2L}$\ decreases smoother
above $M_R$\ and so the unification point with $\a_3$\ is lower again.
As a conclusion we get,at first loop, the upper limit on $M_X^{[6]}$:
$$M_x\leq M_w\exp{{\pi\over2}({\s-{\a\over\a_s}(M_w)\over\a(M_w)})}=2.76\c
10^{15}\c8^{0\pm1}\eqno(2)$$
corresponding to $\tau_p\leq1.6\c10^{33}\cdot8^{0\pm4}$.

The uncertainties depend on the present errors on $\sin^2\theta_w(M_w)$\ and
$\alpha_s(M_w)$.

If the breaking of the $G'$\ symmetry is induced by the VEV of the 126 (and
$\overline{126}$), the Yukawa couplings f$_i$\ of the fermions of the 16
($\overline{16}$) give rise to Majorana masses for the left-handed
antineutrinos of the i-th family given by :
$${M_{\overline\nu_{Li}}=f_i<126>={f_i\over g_{2R}(M_R)}M_R}.\eqno(3)$$
{}From the see-saw mechanism$^{[7]}$\ and (3) we obtain :
$$\eqalignno{M_{\nu_{\tau_L}}&={g_{2R}\over f_3}({10^{11}GeV\over M_R})
({m_t\over100GeV})^2 10eV\cr
M_{\nu_{\mu_L}}&={g_{2R}\over f_2}({10^{11}GeV\over M_R})2\c10^{-3}eV &(4)\cr
M_{\nu_{e_L}}&={g_{2R}\over f_1}({10^{11}GeV\over M_R})2\c10^{-10}eV.\cr}$$
If the spontaneous breaking of $G'$\ is induced by the scalars of the 16 (and
$\overline{16}$)
which cannot have Yukawa couplings to the fermions, one predicts Majorana
masses
for the left-handed antineutrinos (neutrinos) smaller (larger) by several
orders
of magnitude$^{[8]}$.
In Table 2, for the models with $SU(2)_R\subset G'$, we report the values of
$\tau_p$\ and $\mu={10^{11}GeV\over M_R}10eV$\ deduced evaluating $M_X$\ and
$M_R$\ from the renormalization group equations at first (in brakets) and
second loop approximation with the contributions of the scalar multiplets
required by symmetry (the multiplet under $G'$\ containing $<126>$\ above $M_R$
and the electroweak Higgs above $M_w$). For the last possibility in Table 2 we
have taken for $\theta$\ the value chosen in ref.[4]; however, also different
values for $\y$\ have been considered$^{[9]}$\ and the corresponding values of
$M_X$\ $(M_R)$\ may at most increase (decrease) by $10\%$\ $(50\%)$.

As we can see the present uncertainty on the values of  $\s$
and  $\alpha_s(M_w)$\ implies a big uncertainty for the predicted proton
lifetime. Nevertheless we can draw some conclusions; the model with $SU(4)\o D
\subset G'$\ appears, at the second loop approximation, inconsistent with the
experimental lower bound
$${(1-30)\c10^{31}years\leq\tau(p\rightarrow e^+\pi^0)}.\eqno(5)$$
(in all these models, with $SU(2)_R\subset G'$, $Br(p\rightarrow e^+\pi^0)\sim
30\%$.)
The model with $SU(3)\o U(1)\o D\subset G'$\ is consistent with (5) only if
 $\s $\ and/or $\alpha_s$\ are larger than the central values in (1).
Finally the models with $D\not\subset G'$\ are consistent with (5), especially
the one with $SU(3)\otimes U(1)\otimes G'$, which, at first loop, predicts for
$M_X$\ the upper limit in (2).
The present error on $\a_s$\ has twice the effect of the error on
$\s $\ on the determination of $M_X$\ . The e$^{+}$\ e$^{-}$\
experiments will soon give a more precise determination of $\s $;
therefore we study the relationship between  $\s $, $M_R$, and
$M_X$\ in the various models. For $G'=SU(3)\o SU(2)\o SU(2)\o U(1)$\ one has at
first loop :
$${\ln{M_R\over M_w}={\pi\over\a}({3\over8}-\s){16\over17}-{19\over17}
\ln{M_x\over M_w}}\eqno(6)$$
if we define $M^0_x$\ the value corresponding to the lower limit for $\tau_p$\
one gets from (6) the upper limit for $M_R$:
$${M_R\leq M_w({M_w\over M^0_x})^{19\over17}\exp{[{\pi\over\a}({3\over8}-\s)
{16\over17}}]}.\eqno(7)$$
In this way we get for the different models a lower limit for $\mu$\ as a
function of the lower limit on $\tau_p$\ (see Table 3).
\medskip
In conclusion we see that the lower limit on $\mu$\ is an increasing function
of
the lower limit on $\tau_p$. If the intermediate symmetry $G'$\ is broken by
the VEV of the 16, higher values are expected for the masses of the left-handed
neutrinos.
\vfill\eject
\centerline{\bf REFERENCES}
\vskip 0.5cm
[1] U.Amaldi, A. Bohm, L.S. Durkin, P. Langacker, A.K. Mann, W.J. Marciano,
    A. Sirlin and H.H. Williams, Phys. Rev. D  36 (1987) 1385
\medskip
[2] H. Fritzsch and P. Minkowski, Ann. Phys. (NY) 93 (1975) 183
    H. Georgi ``{\ Particles and Fields}'', ed. C.E. Carlson, Am. J. Phys.
    (1975)
\medskip
[3] F. Buccella in ``{\ Symmetry in science III }''eds. B. Gruber and F.
    Iachello (1988)
\medskip
[4] M. Abud, F. Buccella, L.Rosa, and A. Sciarrino, Z. Phys. C -Particles
    and Fields- 44 (1989) 589 and references quoted therein
\medskip
[5] V. Kuzmin and N.Shaposhnikov, Phys. Lett. B 92 (1980) 115;
    D. Chang, R.N. Mohapatra and M.K. Parida, Phys. Lett. B 142 (1984) 55;
Phys.
    Rev. Lett. 52 (1984) 1072; Phys. Rev. D 30 (1984) 1052
\medskip
[6] F. Buccella, G. Miele, L. Rosa, P. Santorelli, and T. Tuzi Phys. Lett. B
    233 (1989) 178
\medskip
[7] M. Gell-Mann, P.Ramond, and R. Slansky  ``{\ Supergravity}''  ed. D.Z.
    Freedman and P. van Nieuwenhuizen (1980)
\medskip
[8] E. Witten Phys. Lett. B 91 (1980) 81
\medskip
[9] G. Amelino-Camelia  thesis (1990)

\end


\def\c{\cdot}
\def\a{\alpha}
\def\ot{\otimes}

\def\ov{\over}

\def\'{\`}
\def\s{\sin^2\theta_W}
\documentstyle [12pt]{report}
\textwidth=18cm
\textheight=23cm
\oddsidemargin=-18mm
\topmargin=-1cm

\begin{document}
\baselineskip=1cm
\scriptsize
\vspace{5mm}
\centerline{\bf TABLE 2}

\smallskip

\begin{center}
\begin{tabular}{|c|c|c|c|c|c|} \hline
 & &\multicolumn{4}{c|}{$SU(2)_L\ot SU(2)_R\ot$}\\ \cline{3-6}
 & &$\ot SU(4)\ot D$& $\ot SU(4)$&$\ot SU{3}\ot U(1)\ot D$&$\ot SU(3)\ot
U(1)$\\ \hline
 & & (I) II loop & (I) II loop & (I) II loop & (I) II loop\\ \hline
 $\s=.232$&$\tau_P\ov10^{31}years $ &(2.9) .12 &$(5.2\cdot10^3)$ 98 &
$(1.2\cdot10^3)$ 5.5 &$(2.3\cdot10^6)$ $8.8\cdot10^2$ \\ \cline{2-6}
 $\a_s=.12$&$\mu\over1eV$ &(.023) .017 &(.85) 1.01 &(28) 3.1 &(310) 20 \\
\hline
$\s=.228$&$\tau_P\ov10^{31}years $&(.041) $3.5\cdot10^{-3}$ &(3.9) .43 &
(2.0) .065 &(160) 2.1 \\ \cline{2-6}
$\a_s=.107$&$\mu\over1eV$ &(.020) .023 &(.25) .43 &(2.4) .96 &(11) 3.6 \\
\hline
$\s=.224$&$\tau_P\ov10^{31}years $&$(9.7\cdot10^{-4})\  1.1\cdot10^{-4}$
&$(6.4\cdot10^{-3})\  3.6\cdot10^{-4}$ &$(6.9\cdot10^{-3})\  3.5\cdot10^{-4}$
&(.029) $5.3\cdot10^{-4}$  \\ \cline{2-6}
$\a_s=.098$&$\mu\over1eV$ &(.016) .019 &(.082) .039 &(.25) .048 &(.55) .068 \\
\hline
\end{tabular}
\vskip 3cm
\centerline{ TABLE 3}

\smallskip
\vskip 0.5cm
\small
\begin{tabular}{|c|c|c|c|c|} \hline
 & &\multicolumn{3}{c|}{$SU(2)_L\ot SU(2)_R\ot$}\\ \cline{3-5}
 & & {$\ot SU(4)$} &{$\ot SU{3}\ot U(1)\ot D$}&{$\ot SU(3)\ot U(1)$}\\ \hline
$\tau_P\geq3\c10^{30}$& ${\mu\over1eV}\geq$& $.19\c3^{0\pm1}$&
$1.4\c4^{0\pm1}$ &
$2.0\c4^{0\pm1}$\\ \hline
$\tau_P\geq9\c10^{31}$& ${\mu\over1eV}\geq$& $.28\c3^{0\pm1}$&
$3.6\c4^{0\pm1}$&
$5.1\c4^{0\pm1}$\\ \hline
\end{tabular}\end{center}
\end{document}